\documentclass{elsart5p}

\usepackage{amssymb,amsmath} 
\usepackage{bm,hyphenat,xspace} %,draftcopy}
\usepackage{graphicx,epsfig}
\usepackage{tabularx}
\newcommand{\Cx}{{}^{18}\mathrm{C}}
\newcommand{\Cn}{{}^{19}\mathrm{C}}
\newcommand{\Cnn}{{}^{20}\mathrm{C}}
\newcommand{\fm}{\mathrm{fm}}
\newcommand{\MeV}{\mathrm{MeV}}
\newcommand{\fmi}{\mathrm{fm}^{-1}}

\newcommand{\email}[1]{\ead{#1}}

\begin{document}
%\linenumbers

\begin{frontmatter}

\title {Neutron-${}^{19}\mathrm{C}$ scattering: 
Towards including realistic interactions}
%Pauli-forbiden states and implications for Efimov physics}

\author{A.~Deltuva}
\email{arnoldas.deltuva@tfai.vu.lt}

%\affiliation{
\address{
Institute of Theoretical Physics and Astronomy, 
Vilnius University, Saul\.etekio al. 3, LT-10257 Vilnius, Lithuania}
%Centro de F\'{\i}sica Nuclear da Universidade de Lisboa, 
%P-1649-003 Lisboa, Portugal }

%\received{October 11, 2015}

%\pacs{24.10.-i, 21.45.-v, 25.45.Hi, 25.40.Hs}

\begin{abstract}
Low-energy neutron-${}^{19}$C scattering is studied in the three-body
 $n + n + {}^{18}$C model using a realistic $nn$ potential and
a number of shallow and  deep $n$-${}^{18}$C potentials, the latter
supporting deeply-bound Pauli-forbidden states that are projected out.
Exact Faddeev-type three-body scattering equations for transition operators 
including two- and three-body forces are solved in the momentum-space
partial-wave framework. Phase shift, inelasticity parameter,
and cross sections are calculated.
For the elastic $n$-${}^{19}$C scattering in the $J^\Pi=0^+$ partial wave
the signatures of the Efimov physics, i.e., the
pole in the effective-range expansion and the elastic
cross section minimum, are confirmed for both shallow and deep models,
but with clear quantitative differences between them,
 indicating the importance
of a proper treatment of  deeply-bound Pauli-forbidden states.
In contrast, the inelasticity  parameter  is mostly correlated with
the asymptotic normalization coefficient of the ${}^{19}$C bound state.
Finally, in the regime of very weak ${}^{19}$C binding and
near-threshold (bound or virtual) excited  ${}^{20}$C state the
standard Efimovian behaviour of the  $n$-${}^{19}$C scattering length and
cross section was confirmed, resolving the discrepancies
between earlier studies by other authors
[I. Mazumdar, A.~R.~P. Rau, V.~S. Bhasin, Phys. Rev. Lett. 97  (2006) 062503;
M.~T. Yamashita, T. Frederico, L. Tomio, Phys. Rev. Lett. 99  (2007) 269201].

\end{abstract}


\end{frontmatter}
% \maketitle

%%%%%%%%%%%%%%%%%%%%%%%%%%%%%%%%%%%%%%%%%%%%%%%%%%%%%%%%%%%%%%%%%%%%%%%%%%%%%%%
\section{Introduction \label{sec:intro}}

Few-particle systems whose  two-particle ($ij$) subsystems are characterized
by large $s$-wave scattering lengths $a_{ij}$ exhibit universal properties.
Their systematic theoretical study  was pioneered 
by V. Efimov almost 50 years ago \cite{efimov:plb}
but the experimental confirmation came many years later
\cite{kraemer:06a,pollack:09a,zaccanti:09a,PhysRevLett.103.043201}.
It was achieved in  cold-atom  systems where the two-atom scattering
length in the vicinity of the Feshbach resonance can be controlled by 
an external magnetic field and thereby tuned to a large value significantly
exceeding the interaction range, a condition needed for the manifestation
of the so-called universal or Efimov physics. The possibility of tuning
the scattering length is not available in the nuclear physics. Nevertheless,
some nuclear systems have quite large two-particle scattering lengths 
and qualitatively show some properties characteristic for Efimov physics.
The simplest case is the three-nucleon system
\cite{efimov:plb,PhysRevC.44.2303,Bedaque1998221,Bedaque2000357,Bedaque2003589,Konig2016,PhysRevC.95.024001}.
Further examples are systems consisting of a nuclear core ($A$) and 
two neutrons ($n$)
provided that there is a weakly bound or virtual $s$-wave state in the
$(A+n)$ subsystem
\cite{jensen:04a,braaten:rev,Hammer2010,frederico:12a,hammer:17a}.
Among them the $\Cx + n + n$ system has been studied
in a number of works (see Refs.~\cite{frederico:12a,hammer:17a} for a review)
hoping to establish the existence of a $\Cnn$ excited Efimov state
assuming  that $\Cn$  has the binding energy of only 
$S_{n} = 0.16$ MeV \cite{AUDI2003337}
while the ground state of $\Cnn$ is bound with $S_{2n} = 3.5$ MeV 
(relative to the $\Cx + n + n$ threshold).
However, more recent experiments have not confirmed such a weak
binding of $\Cn$ and presently accepted value is $S_{n} =0.58(9)$ MeV
\cite{ame2012}, 
thereby excluding also the possibility of excited Efimov state in $\Cnn$
as a real bound state. Nevertheless, the $\Cx + n + n$ system in the
low-energy $n+\Cn$ scattering process is expected to exhibit some universal
properties that have been studied theoretically both in zero-range
and finite-range models  
\cite{PhysRevC.69.061301,PhysRevLett.97.062503,Yamashita200849,shalchi:17a}.
However, there is no consensus on the fate of the $\Cnn$ excited Efimov state
as $\Cn$ binding increases towards its physical value.
Refs.~\cite{Yamashita200849,shalchi:17a,PhysRevLett.99.269201} 
predict that it becomes a virtual state leading to a pole
in the effective-range expansion  for $n+\Cn$ scattering 
similar to the neutron-deuteron case \cite{VANOERS1967562} while 
Refs.~\cite{PhysRevC.69.061301,PhysRevLett.97.062503,PhysRevLett.99.269202} 
claim that the  $\Cnn$ excited state turns into
a continuum resonance seen as a pronounced peak in the 
$n+\Cn$ elastic cross section around 1.5 keV center-of-mass (c.m.) energy.
One of the conclusions drawn in Ref.~\cite{PhysRevLett.99.269202}
was that ``there is a need to undertake a
detailed investigation using realistic interaction''.
Indeed, all the calculations for the $n+\Cn$ scattering so far have been
performed using simple rank-one separable potentials with
Yamaguchi form factors for $n$-$n$ and  $n$-$\Cx$ pairs.
Although in the ideally universal regime the predictions for observables
should be independent of the interaction details, some remnant
dependence is expected for realistic systems whose universal behaviour
is modified by finite-range corrections.

The goal of the present work is to study the 
low-energy $n+\Cn$ scattering using more realistic interactions,
possibly establishing shortcomings of rank-one separable models,
and sort out the differences between findings of Refs.
\cite{PhysRevC.69.061301,PhysRevLett.97.062503,PhysRevLett.99.269202}
 and \cite{Yamashita200849,shalchi:17a,PhysRevLett.99.269201}.
The improvement of the interaction models is threefold:
(i) For the $n$-$n$ pair a realistic high-precision
charge-dependent (CD) Bonn potential \cite{machleidt:01a} is used.
(ii) A rank-one separable potential can support at most one
bound state, thus, it misses deeply-bound core-neutron states. These states
are occupied by internal neutrons in the core (not treated explicitly) and
therefore are Pauli-forbidden and must be projected out. In this way
the identity of external neutrons and those
within the core is approximately taken into account.
 A proper treatment of
Pauli-forbidden states was found to be important in scattering
processes, see e.g. Ref.~\cite{deltuva:06b} for 
$\alpha$-deuteron collisions. Using a  $n$-$\Cx$ potential
supporting the $2s$ state with the 0.58 MeV experimental value
\cite{ame2012} of the $\Cn$ binding energy and projecting out 
the deep $1s$ state will enable to study the importance
of the Pauli-forbidden state for the $n+\Cn$ scattering
and its impact on the Efimov physics.
(iii) Depending on the chosen two-particle potentials, an additional
 three-body force (3BF) may be needed to fix the $\Cnn$ ground-state
binding energy that must be included also in the $n+\Cn$ scattering
calculations.

Thus, for the desired study of the $n+\Cn$ scattering  an accurate
theoretical description of three-particle scattering process 
including general form of potentials and 3BF is needed;
the separable  quasi-particle formulation of 
Refs.~\cite{PhysRevC.69.061301,PhysRevLett.97.062503,Yamashita200849,shalchi:17a} 
is not applicable.
In the present work the description is obtained by a combination
and extension of momentum-space 
techniques from Refs.~\cite{deltuva:06b,deltuva:09e}.
Both are based on exact Faddeev three-body theory \cite{faddeev:60a}
in the integral form for transition operators as proposed by Alt,
 Grassberger, and Sandhas (AGS) \cite{alt:67a}, but either neglecting
the 3BF \cite{deltuva:06b} or limited to the three-nucleon system
\cite{deltuva:09e}.

The employed   potentials are described 
in Sec.~\ref{sec:2} and  the three-body scattering equations with 3BF
in Sec.~\ref{sec:3}.
Results are presented in  Sec.~\ref{sec:4},
and a summary is given in  Sec.~\ref{sec:5}.

\section{Potentials \label{sec:2}}

The system of two neutrons and  $\Cx$ core is considered
as a three-body problem.
Particle masses $m_n = 1.00069\,m_N$ and
 $m_A = 18\,m_N$ are given in units of the average nucleon mass
  $m_N = 938.919$ MeV. The dynamics of the system is determined
by two-particle potentials $v_{nn}$ and $v_{nA}$ acting within
the $nn$ pair and two  $nA$ pairs, and, eventually, by an
additional 3BF. Unless explicitly stated otherwise, 
$v_{nn}$ is taken to be the high-precision
CD Bonn potential \cite{machleidt:01a};
it is allowed to act in the $s$, $p$ and $d$ waves thereby
ensuring a perfect convergence of the  $nn$ partial-wave expansion.
Given the uncertainty in the
$v_{nA}$, a number of models will be used. A very common choice
is the Woods-Saxon potential, in the coordinate space defined as
\begin{gather}  \label{eq:WS}
\bar{v}_{nA}(r) =  {} -V_c [1+\exp((r-R_v)/a_v)]^{-1}.
\end{gather}
In the present work this kind of potentials is numerically transformed
into the momentum-space representation and then used in two- and 
three-particle equations. The potentials parameters are
 the strength $V_c$, radius $R_v = r_v A^{1/3}$, and diffuseness $a_v$ with
standard  values being around $r_v = 1.2$ fm and $a_v = 0.6$ fm;
this parameter set is among the considered models. Two further
models are taken as $r_v = 0.8$ fm and $r_v = 0.332$ fm keeping
the ratio $a_v/r_v = 1/2$. In all three cases the strength $V_c$ is 
adjusted such that the excited $s$-wave state $2s$
has the experimental
$\Cn$ binding energy value $\varepsilon_{2s} = S_n = 0.58$ MeV.
The ground state is deeply bound with the energy 
 $\varepsilon_{1s}$ and the wave function $|\phi_{1s}\rangle$ 
 depending on the chosen  $r_v$,  
but it is Pauli-forbidden and therefore must be projected out. 
This is achieved \cite{schellingerhout:93a,thompson:00,deltuva:06b}
by taking the neutron-core $s$-wave potential as
\begin{gather}
v_{nA}^s = \bar{v}_{nA} + |\phi_{1s}\rangle \Gamma \langle \phi_{1s}|
\end{gather}
where formally $\Gamma \to \infty$ but in practice $\Gamma$ must
be large enough such that the results for the three-body bound state(s) and
scattering in the considered energy regime become independent of it;
in the present work  $\Gamma = 50$ GeV was proven to be sufficiently large.
Simultaneously this ensures also the absence of deeply-bound three-body states.
The potential models projecting out deeply-bound Pauli-forbidden (DP)
states will be denoted as DPa, DPb, and DPc; their parameters
are collected in table \ref{tab:Vd}. The predictions for the $nA$ scattering
length $a_{nA}$ and  the effective range  $r_{nA}$ are presented as well.
As the potentials of
Refs.~\cite{PhysRevC.69.061301,PhysRevLett.97.062503,Yamashita200849,shalchi:17a},
 they are restricted to act in the $s$-wave only.
The manifestation of the Efimov physics is governed by resonant
$s$-wave interactions but in realistic systems also higher
partial waves contribute. To estimate their effect, one more model,
labeled DPp, is introduced that uses $\bar{v}_{nA}$ parameters
from DPa but is allowed to act in $p$-waves as well. It turns out that
$\bar{v}_{nA}$ supports a deeply-bound $1p$ state $|\phi_{1p}\rangle$ for $\Cn$ with
$\varepsilon_{1p} = 11.54$ MeV that is Pauli-forbidden as well and must 
be projected out in the same way, i.e.,
\begin{gather}
v_{nA}^p = \bar{v}_{nA} + |\phi_{1p}\rangle \Gamma \langle \phi_{1p}|.
\end{gather}

Furthermore, as in Refs.~\cite{Yamashita200849,shalchi:17a} the energy 
of the three-body bound state, i.e., the two-neutron separation energy 
of  $\Cnn$, is fixed at its experimental value of $S_{2n} = 3.5$ MeV. 
Except for the DPc model whose range $R_v$ was adjusted to reproduce $S_{2n}$,
in general case
the pairwise $nn$ and $nA$ interactions are insufficient for $S_{2n}$ and 
an additional 3BF is needed. In fact, when the 3BF is not included,
$S_{2n} = 2.083$, 2.404, and 2.126 MeV for DPa, DPb, and DPp, respectively.
Being unable to derive the 3BF from a microscopic
many-nucleon theory, usually a phenomenological form of the 3BF is assumed,
depending either on the hyperradius \cite{thompson:00} or 
hypermomentum \cite{hammer:07a,deltuva:13b}. The latter choice is obviously more
convenient in the momentum-space framework and therefore is used in the present
calculations. The three-body bound-state Faddeev equations,
their solution technique, 
and the form of the 3BF is taken over from Ref.~\cite{deltuva:13b}. The latter is
\begin{equation} \label{eq:3bf}
\langle  \mathbf{p}_\alpha \mathbf{q}_\alpha 
| W | \mathbf{p}'_\alpha \mathbf{q}'_\alpha   \rangle = 
-(4\pi)^{-2} W_c g(\mathcal{K}^2) g(\mathcal{{K}'}^2)
\end{equation}
with  the hypermomentum
$\mathcal{K}^2 = m_N(p_\alpha^2/\mu_\alpha+ q_\alpha^2/M_\alpha)$ 
expressed in terms of Jacobi momenta  
$\mathbf{p}_\alpha$ for the pair and $\mathbf{q}_\alpha$ for the spectator
  and the associated reduced masses $\mu_\alpha$ and $M_\alpha$.
 Note that $\mathcal{K}^2/2m_N$ is the internal motion kinetic energy,
thus, the 3BF has the same form in any Jacobi configuration labeled by the
spectator particle $\alpha$ in the odd-man-out notation 
(see next section for more details).
The form factor 
$g(\mathcal{K}^2) = \exp{(-\mathcal{K}^2/2\Lambda^2)}$ is chosen as a Gaussian.
The cutoff parameter $\Lambda$ is related to the interaction range $R_w$ roughly
as $\Lambda R_w \sim \sqrt{2}$; for each two-body model 
$R_w/R_v \sim \sqrt{2}/(\Lambda R_v) < 1/2$ ensuring that the 3BF is 
of shorter range than $v_{nA}$. The  strength  $W_c$ is adjusted
to reproduce the desired three-body binding energy. These parameters are
collected in table \ref{tab:Vd} as well. 

To isolate the effect of deeply-bound Pauli-forbidden states, several
models without those states are used, i.e., they support only one
$\Cn$ bound state $1s$ with the binding energy 
$\varepsilon_{1s} = 0.58$ MeV. The model WS uses a shallow
 Woods-Saxon potential  (\ref{eq:WS}) with $r_v = 1.19$ fm, while
the model labeled Y is a rank-one separable potential with Yamaguchi
form factor
as in Ref.~\cite{shalchi:17a} with the momentum-range parameter
$\beta_{nc} = 0.6167\, \fmi$. Finally, to get the insight
on the importance of a realistic $nn$ interaction, instead of the
CD Bonn the rank-one separable $nn$ potential with Yamaguchi
form factor from Ref.~\cite{shalchi:17a} is used;
its combination with the $nA$ potential of the same type as Y but
with $\beta_{nc} = 0.6131\, \fmi$ will be labeled YY
in the following. The above choices of the range parameter values
for WS, Y, and YY models ensure the desired binding energy of
$\Cnn$ without the 3BF. The WS, Y, and YY potentials
will be referred in the following as {\it shallow}, in contrast to
the {\it deep} ones DPa, DPb, DPc, and DPp.

\begin{table} [!]
\caption{\label{tab:Vd}
Parameters of the employed  $nA$ force models DPa, DPb, DPc and DPp
with projected-out deeply-bound Pauli-forbidden states 
and 3BF. In the last line the parameters of the shallow WS 
potential are given.}
%\begin{ruledtabular}
\setlength{\tabcolsep}{5pt}
\begin{tabularx}{\columnwidth}{l|*{2}{r}c*{2}{r}}
\hline
 & $R_v$(fm) &  $V_c$(MeV) & $\Lambda(\fmi)$ & $W_c(\fm^6 \MeV)$ 
& $\varepsilon_{1s}$(MeV)  \\ \hline
 DPa & 3.145 & 44.569 & 1.0  & 312.40 & 25.52 \\
 DPb & 2.097 & 95.995 & 1.5  & 43.34 & 53.98 \\
 DPc & 0.870 & 530.745 &   &  0.00 & 292.00 \\
 DPp & 3.145 & 44.569 & 1.0  & 287.50 & 25.52 \\ \hline
 WS &  3.119 & 8.317  &      & 0.00   & 0.58 \\
\hline
\end{tabularx}
%\end{ruledtabular}
\end{table}

To get an insight into the correlations between the
interaction models and physical properties of $\Cn$ and $\Cnn$,
in table \ref{tab:Vpred} the predictions  for the 
$n$-$\Cx$ scattering length  $a_{nA}$ and effective range  $r_{nA}$,
the asymptotic normalization coefficient (ANC) of the $\Cn$ bound state, 
and the $\Cnn$ ground state internal kinetic energy expectation 
value $\bar{K}_{3b}$ are collected. Within the group of deep models
one may easily notice the well-known
feature that a longer-range potential leads to larger values of the ANC, 
effective range, and, to a lesser extent, scattering length. However,
comparing DPa and WS that have almost the same $R$, one can conclude
that deeply-bound Pauli-forbidden states cause 
larger $a_{nA}$,  $r_{nA}$, and ANC values, and significantly higher
 $\bar{K}_{3b}$. Within the  group of DP models the kinetic energy expectation 
value depends also strongly on the range $R$, but in all cases 
it considerably exceeds the predictions of shallow potentials.
Thus, deep potentials strongly enhance high-momentum components
in the $\Cnn$ ground state.

\begin{table} [!]
\caption{\label{tab:Vpred}
Predictions of the employed force models for the 
$n$-$\Cx$ scattering length  $a_{nA}$ and effective range  $r_{nA}$,
ANC of the $\Cn$ nucleus, 
and the internal kinetic energy expectation value $\bar{K}_{3b}$ of the
$\Cnn$ nucleus with the binding energy of $S_{2n} =3.5$ MeV.}
%\begin{ruledtabular}
\setlength{\tabcolsep}{5pt}
\begin{tabularx}{\columnwidth}{l|*{4}{c}}
\hline
& $a_{nA}$(fm) & $r_{nA}$(fm) & ANC$(\fm^{-1/2})$ & $\bar{K}_{3b}$ (MeV) 
\\ \hline
 DPa & 9.23 & 4.32 & 0.948 & 32.83 \\
 DPb & 8.22 & 3.19 & 0.802 & 47.00 \\
 DPc & 7.01 & 1.54 & 0.657 & 86.82 \\  
 DPp & 9.23 & 4.32 & 0.948 & 33.12 \\ \hline
 WS  & 8.12 & 3.04 & 0.792 & 9.87 \\
 Y   & 8.67 & 3.66 & 0.871 & 10.64 \\ 
 YY  & 8.68 & 3.67 & 0.872 & 11.14 \\
\hline
\end{tabularx}
%\end{ruledtabular}
\end{table}

\section{Neutron-$\Cn$ scattering equations including 3BF
\label{sec:3}}

The momentum-space formulation of the three-body scattering
theory is convenient when the underlying potentials have nonlocal
terms such as those in the deep $nA$ potentials  projecting out
Pauli-forbidden states. The present work is based on  Faddeev equations
for the multichannel transition operators $U_{\beta \alpha}$
in the version derived by  Alt, Grassberger, 
and Sandhas (AGS) \cite{alt:67a}  but  extended to include 
also the 3BF. Such extensions have been proposed in a number of works,
e.g., \cite{gloeckle:96a,deltuva:09e}, but their practical applications
mostly were limited to the symmetrized version in the
three-nucleon system. The general form of three-body equations from
Ref.~\cite{deltuva:09e} is taken for the
present study of the $n$-$\Cn$ scattering, i.e., 
\begin{gather} \label{eq:Uba}
\begin{split}
 U_{\beta \alpha} = {} & \bar{\delta}_{\beta \alpha} G_0^{-1} + u_\alpha
+ \sum_{\gamma=1}^3  \bar{\delta}_{\beta \gamma} T_\gamma G_0 U_{\gamma \alpha} \\
 & + \sum_{\gamma=1}^3  u_\gamma G_0(1+T_\gamma G_0)  U_{\gamma \alpha},
\end{split}
\end{gather}
with $\bar{\delta}_{\beta \alpha} = 1 - {\delta}_{\beta \alpha}$,
the free resolvent $G_0 = (E+i0 - H_0)^{-1}$ at the available energy $E$,  
the free Hamiltonian for the internal motion $H_0$,
the two-body transition matrix
\begin{equation}  \label{eq:T}
 T_\gamma =  v_\gamma + v_\gamma G_0 T_\gamma,
\end{equation}
and the 3BF arbitrarily decomposed into three components
\begin{equation} \label{eq:Wu}
 W = \sum_{\alpha=1}^3 u_\alpha.
\end{equation}
The odd-man-out notation is used, i.e., the channel $\alpha$
corresponds to the configuration where the particle $\alpha$
is the spectator and the remaining two are the pair.
The decomposition of the 3BF into three symmetric parts 
(\ref{eq:Wu}) is essential for the symmetrization of three-nucleon
equations \cite{deltuva:09e} but is not needed in the present work.
Labeling the particles $n,A,n$ as $1,2,3$ and taking 
 $ u_\alpha = {\delta}_{\alpha 2} W$, the system of the AGS equations
(\ref{eq:Uba}) for the $n$-$\Cn$ scattering simplifies  to
\begin{gather} \label{eq:Ub1}
%\begin{split}
 U_{\beta 1} = {}  \bar{\delta}_{\beta 1} G_0^{-1}
+ \sum_{\gamma=1}^3  \bar{\delta}_{\beta \gamma} T_\gamma G_0 U_{\gamma 1} %\\ &
  + W G_0(1+T_2 G_0)  U_{21}
%\end{split}
\end{gather}
with $\beta = 1,2,3$. The above system of integral equations is
 solved  in the momentum-space partial-wave representation
employing three  sets   of base functions
$|p_\alpha q_\alpha 
(l_\alpha \{ [ L_\alpha (s_\beta s_\gamma)S_\alpha] j_\alpha s_\alpha\} \mathcal
{S}_\alpha) J M \rangle $ 
with $(\alpha,\beta,\gamma)$ being cyclic permutations of
$(1,2,3)$.
Here $p_\alpha$ and  $q_\alpha$ are magnitudes of Jacobi momenta
for the corresponding pair and spectator, while
 $L_\alpha$ and  $l_\alpha$ are the associated orbital
angular momenta, respectively. Together with the particle spins
 $s_\alpha,s_\beta, s_\gamma$  they are coupled, through the 
 intermediate subsystem spins $S_\alpha$, $j_\alpha$ and $ \mathcal{S}_\alpha$,
 to the total angular momentum $J$ with the projection $M$.
Only the basis $\alpha=2$ is antisymmetric with respect to 
the permutation of the two neutrons; this is achieved by 
considering only even $L_2 + S_2$ states. However, the neutron identity
is accounted for by taking the  antisymmetrized
elastic scattering amplitude
\begin{gather} \label{eq:fasym}
\begin{split}
 f_{\nu'\nu}(\mathbf{k}',\mathbf{k}) = {} & -(2\pi)^2 M_1 
[ \langle \Phi_1^{\nu'}(\mathbf{k}') |U_{11}|\Phi_1^{\nu}(\mathbf{k}) \rangle 
\\ & 
-\langle \Phi_3^{\nu'}(\mathbf{k}') |U_{31}|\Phi_1^{\nu}(\mathbf{k}) \rangle ].
\end{split}
\end{gather}
Here $| \Phi_\alpha^{\nu} (\mathbf{k}) \rangle$ is the asymptotic state
in the channel $\alpha$;
it is given by the product of the  bound state wave function for the pair
and the plane wave with the on-shell momentum  $\mathbf{k}$
for the relative motion between the bound pair and
 spectator $\alpha$ satisfying $E = -S_n +  k^2/2M_1$;  
the spin quantum numbers are abbreviated by ${\nu}$.
 In the normalization of Eq.~(\ref{eq:fasym})
the $n$-$\Cn$ elastic differential cross section for the 
$\nu\mathbf{k} \to \nu'\mathbf{k}'$ transition  is simply
$d\sigma/d\Omega = |f_{\nu'\nu}(\mathbf{k}',\mathbf{k})|^2$.

\section{Results \label{sec:4}}

The Efimov physics manifests itself in the states dominated by
the $s$-wave components $L_\alpha = l_\alpha =0$ for all $\alpha$;
this condition is satisfied only for $J^\Pi=0^+$ where
$\Pi=(-1)^{L_\alpha + l_\alpha}$ is the total parity. For the notational
 brevity suppressing the dependence on the on-shell momentum $k$,
the $S$-matrix and
the amplitude in the  $0^+$ state are parametrized as
$s = e^{2i\delta}$ and 
$f = e^{i\delta} \sin{\delta}/k = (k\cot{\delta}-ik)^{-1}$,
respectively. The phase shift $\delta$ is real below the
inelastic threshold, i.e., at c.m. kinetic energies
$E_k = k^2/2M_1 \le 0.58$ MeV, but becomes complex above this value
due to the open breakup channel whose importance is parametrized
by the inelasticity parameter $\eta = |e^{2i\delta}| \le 1$. 

\begin{table} [!]
\caption{\label{tab:abc}
Parameters of the $n$-$\Cn$ effective-range expansion
 for  the employed interaction models
together with $r_{nA}$ for $n$-$\Cx$.}
%\begin{ruledtabular}
\setlength{\tabcolsep}{3pt}
\begin{tabularx}{\columnwidth}{l|*{5}{c}}
\hline
& $a$(fm) & $b(\fm^{-1}\,\MeV^{-1})$ & $c(\fm^{-1}\,\MeV^{-2})$ & $E_{0}$(MeV) & $r_{nA}$(fm) 
\\ \hline
 DPa & -6.299 & 1.176 & -0.2726 & 0.20626 & 4.32 \\
 DPb & -6.103 & 1.078 & -0.1538 & 0.26235 & 3.19 \\
 DPc & -5.369 & 1.033 & -0.0511 & 0.33770 & 1.54 \\  
 DPp & -6.310 & 1.176 & -0.2744 & 0.20644 & 4.32 \\
\hline
 WS  & -9.419 & 0.8194 & -0.0496 & 0.39663 & 3.04 \\
 Y   & -9.802 & 0.8520 & -0.0838 & 0.34710 & 3.66 \\ 
 YY  & -9.653 & 0.8591 & -0.0828 & 0.34413 & 3.67 \\
\hline
\end{tabularx}
%\end{ruledtabular}
\end{table}

\begin{figure}[!]
\begin{center}
\includegraphics[scale=0.69]{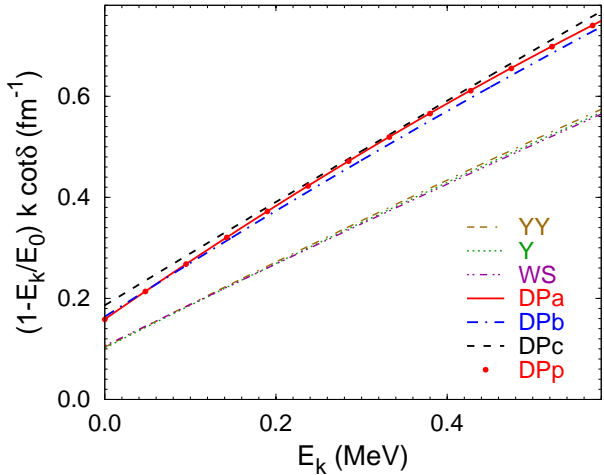}
\end{center}
\caption{\label{fig:kctg}  (Color online)
$n$-$\Cn$ reduced effective-range functions $(1- E_k/E_0)k\cot{\delta}$ 
in the $J^\Pi=0^+$  partial wave for
the interaction models YY(double-dashed-dotted), Y (dotted), 
WS (double-dotted-dashed), DPa (solid), DPb (dashed-dotted),
DPc (dashed), and DPp (bullets).}
\end{figure}

The presence of the virtual Efimov state leads to a modified
 effective range expansion \cite{VANOERS1967562,shalchi:17a} 
containing a pole, i.e.,
\begin{equation} \label{eq:kctg}
k\cot{\delta} \approx \frac{-a^{-1} + b E_k + c E_k^2}{1- E_k/E_0},
\end{equation}
where $a$ is the $n$-$\Cn$ singlet scattering length and $E_0$ is the position
of the pole. The values for the parameters $a$, $b$, $c$, and $E_0$
 obtained fitting the $n$-$\Cn$ phase shift results 
at $E_k \le 0.58$ MeV for all employed interaction models are
collected in table \ref{tab:abc} while the corresponding reduced
effective-range functions
$(1- E_k/E_0)k\cot{\delta}$ are plotted in Fig.~\ref{fig:kctg}.
It turns out that Eq.~(\ref{eq:kctg}) yields a very good approximation
- the quantities calculated directly and from the fitted parameters
are indistinguishable in the plot. One notices immediately that
$(1- E_k/E_0)k\cot{\delta}$ predictions for the groups of 
the shallow (YY,Y,WS) and deep (DPa,DPb,DPc,DPp) potentials clearly
separate. A closer inspection of the table \ref{tab:abc} reveals
that this is mostly due to the differences in the 
$n$-$\Cn$ scattering length $a$ and, to a lesser extent, in the parameter
$b$. Within each group one can see qualitatively the same trend
in correlations between the $n$-$\Cx$ effective range $r_{nA}$ and
$n$-$\Cn$ parameters, i.e., $|a|$, $b$ and $|c|$ increase with
increasing  $r_{nA}$ while $E_0$ decreases. However, it turns out that
the presence of deep Pauli-forbidden states is more decisive for $a$ and $b$
than the correlation with $r_{nA}$, while for $c$ and $E_0$ 
these two effects are of comparable importance.
 The  parameters $c$ and $E_0$ show a broad
spread of values, especially in the group of deep potentials. However,
if one disregards the DPc model as being of unrealistically short range,
one can see again some trend, i.e., larger $|c|$ and smaller $E_0$
for deep potentials as compared to  shallow ones. The parameters
of DPa and DPp stay very close indicating that the $n$-$\Cx$ $p$-wave 
interaction is indeed irrelevant in the present context.
The deviations  between Y and YY for all parameters are insignificant
 as well, thus, the rank-one separable $s$-wave $nn$ potential 
is able to capture relevant physics for the $n$-$\Cn$ low-energy 
$J^\Pi=0^+$ elastic scattering.

\begin{figure}[!]
\begin{center}
\includegraphics[scale=0.69]{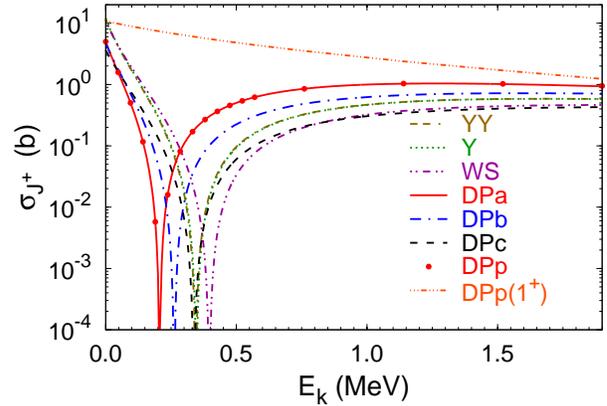}
\end{center}
\caption{\label{fig:csel}  (Color online)
 $n$-$\Cn$ total elastic cross section $\sigma_{0^+}$ in the
$J^\Pi=0^+$  partial wave as a function of the c.m. kinetic energy $E_k$
for different interaction models. In addition, the
$J^\Pi=1^+$ wave cross section $\sigma_{1^+}$ is shown for the
DPp model as the upper triple-dotted-dashed curve,
other curves are as in Fig.~\ref{fig:kctg}.}
\end{figure}

The differences in $a$ and $E_0$ are clearly reflected in the $J^\Pi=0^+$
total elastic cross section $\sigma_{0^+}$ for the $n$-$\Cn$ 
scattering shown in Fig.~\ref{fig:csel}: $a$ determines
$\sigma_{0^+}$ at  $E_k = 0$ while  $E_k=E_0$ corresponds to the 
minimum of $\sigma_{0^+}$. However, this minimum is only
clearly seen when the initial state $n$ and $\Cn$ spins
are anti-parallel, such that the total channel spin
$ \mathcal{S}_1 = 0$ couples with $l_1 = 0$ to $J^\Pi=0^+$.
If the initial state is not polarized, one has to take into account
also the  $n$+$\Cn$ triplet state $(\mathcal{S}_1 = 1, l_1 = 0)J^\Pi=1^+$
whose cross section  $\sigma_{1^+}$ is also shown in  Fig.~\ref{fig:csel}.
In fact, $\sigma_{1^+}$ yields by far the most sizable contribution to
the unpolarized low-energy cross section, given (neglecting 
$l_1=1$ and higher waves) as the spin-weighted average
$ \sigma = (\sigma_{0^+} +3\sigma_{1^+})/4$. Since the ${}^1S_0(nn)$
configuration  is not allowed in the $J^\Pi=1^+$ state, $\sigma_{1^+}$
is governed by the $nA$ interaction. In fact, for all models
with the $nn$ CD Bonn potential  the $n$-$\Cn$ triplet scattering length
$a_{1^+} \approx a_{nA} + 0.02\, \fm$ is simply related to the
$n$-$\Cx$  scattering length.
 
The results  in  Fig.~\ref{fig:csel} extend above the breakup threshold;
in that regime $\sigma_{0^+}$ depends on $E_k$ only weakly,
with the deep models (except for DPc) providing higher cross section
than the shallow ones, although the spread within each group is 
comparable to the difference between groups. The DPa-DPp and
Y-YY similarities remain valid also over the broader regime.

\begin{figure}[!]
\begin{center}
\includegraphics[scale=0.6]{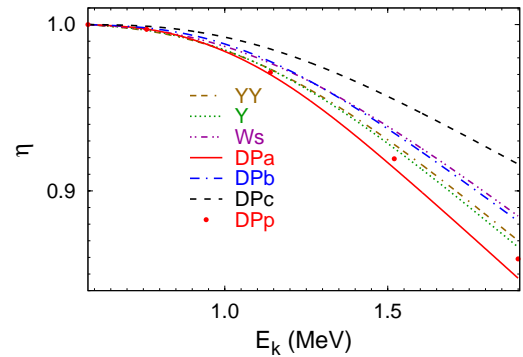}
\end{center}
\caption{\label{fig:inel}  (Color online)
 $n$-$\Cn$ inelasticity parameter $\eta$ in the
$J^\Pi=0^+$  partial wave as a function of the c.m. kinetic energy $E_k$
for different interaction models.
Curves are as in Fig.~\ref{fig:kctg}.}
\end{figure}

\begin{figure}[!]
\begin{center}
\includegraphics[scale=0.6]{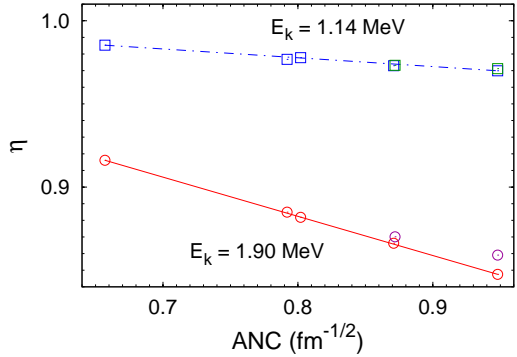}
\end{center}
\caption{\label{fig:inel-anc}  (Color online)
 $n$-$\Cn$ inelasticity parameter $\eta$ in the
$J^\Pi=0^+$  partial wave at  $E_k = 1.14$ (boxes) and 1.90 MeV (circles).
The symbols from left to right 
correspond to the interaction models DPc, WS, DPb, YY, Y, DPa, and DPp.
The lines are for guiding the eye only.}
\end{figure}

However, the situation is quite different for the
inelasticity parameter  $\eta$ studied in 
Fig.~\ref{fig:inel}. It exhibits some DPa-DPp and Y-YY deviations
but shows no trend for the differences between shallow and deep
potentials, the spread for the latter being very broad.
Looking back to the model properties  in table 
\ref{tab:Vpred}, one may notice the correlations 
between the ANC (or $a_{nA}$, or $r_{nA}$) and   $\eta$. 
To make it more evident, the inelasticity parameter 
at $E_k = 1.14$ and 1.90 MeV for all force models is plotted
in  Fig.~\ref{fig:inel-anc} against the corresponding ANC value.
The dependence is roughly linear with deviations by YY and DPp
models, that either have a different $nn$ force (YY) 
from all the others, or have an additional $nA$ $p$-wave dynamics
(DPp). This is not surprising since one can expect the $n$-$\Cn$ breakup
reaction to be peripheral at these low energies and
 dominated by the mechanism of  the $n$-$n$ knockout.
In fact, even neglecting the 3BF for DPa, DPb and DPp models leads to changes 
of $\eta$ that are significantly smaller than the spread of predictions 
in  Fig.~\ref{fig:inel}.
Thus, the breakup and inelasticity parameter  $\eta$ is mostly governed
by the properties  of the $\Cn$ bound state and the $nn$ force,
i.e., by the two-body physics without clear evidence for the
three-body Efimov physics.

\begin{figure}[!]
\begin{center}
\includegraphics[scale=0.66]{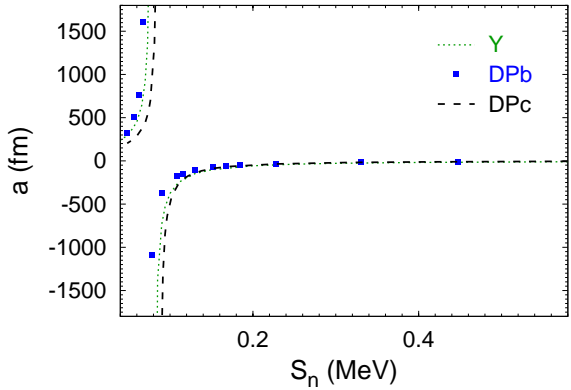}
\end{center}
\caption{\label{fig:a-b}  (Color online)
Dependence of the $n$-$\Cn$ singlet scattering length $a$ on the 
$\Cn$ binding energy $S_n$ for the evolved models
Y (dotted),  DPb (filled boxes) and DPc (dashed).}
\end{figure}

\begin{figure}[!]
\begin{center}
\includegraphics[scale=0.66]{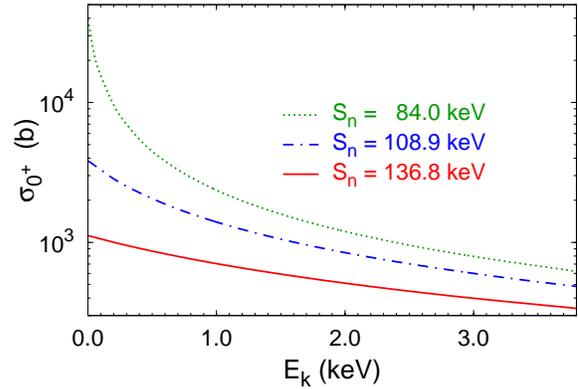}
\end{center}
\caption{\label{fig:cs-nores}  (Color online)
 $n$-$\Cn$ total elastic cross section $\sigma_{0^+}$ in the
$J^\Pi=0^+$ partial wave as a function of the c.m. kinetic energy $E_k$
for  the DPb model  evolved to have the $\Cn$ binding energy $S_n$
of 84.0 keV (dotted), 108.9 keV (dashed-dotted) and 
136.8 keV (solid). The $\Cnn$ bound excited state exists
below $S_n = 71.66$ keV.}
\end{figure}

Finally I turn to the  disagreement between Refs.
\cite{PhysRevC.69.061301,PhysRevLett.97.062503,PhysRevLett.99.269202}
 and \cite{Yamashita200849,shalchi:17a,PhysRevLett.99.269201} near the 
regime  where the bound excited $\Cnn$ Efimov state disappears.
To reach that regime the strength of the $nA$ potential $V_c$ is reduced
without changes in other force parameters; this leads to the variations 
of $\Cn$ and  $\Cnn$ binding energies and $n$-$\Cn$ scattering observables.
The appearance of the bound excited $\Cnn$ state, 
depending on the potential, takes place
when $S_n$ is reduced to 0.07 - 0.09 MeV and $S_{2n}$ to 1.4 - 1.9 MeV.
This is different from the strategy of Ref.~\cite{shalchi:17a}
where $S_{2n}$ was fixed at 3.5 MeV, but nevertheless the present
results support the conclusions of Refs.~\cite{Yamashita200849,shalchi:17a}
that the excited $\Cnn$ state at the $n$+$\Cn$ threshold
corresponds to a pole in the  $n$-$\Cn$ scattering length, i.e.,
 $a \to \pm \infty$, with $+(-)$ for $S_n$ below (above) the critical value.
 This behaviour is shown in Fig.~\ref{fig:a-b}
for selected potential models, but is characteristic for all of them.
 In contrast, the authors of
Refs.~\cite{PhysRevC.69.061301,PhysRevLett.97.062503,PhysRevLett.99.269202}
claim that the  $n$-$\Cn$ scattering length remains positive also
for  $S_n$ above the critical value  while the low-energy elastic 
$n$-$\Cn$  cross section exhibits a resonance
 around $E_k = 1.5$ keV on top of a nearly constant background.
 A thorough study of the $n$-$\Cn$ scattering
in this regime performed in the present work excludes such a behaviour:
the  cross section rapidly and monotonically  decreases with increasing
 energy without any signs of resonant peaks. As example the  $J^\Pi=0^+$
elastic cross section calculated using the evolved DPb model is shown
in Fig.~\ref{fig:cs-nores}.

\section{Summary and conclusions \label{sec:5}}

Low-energy neutron-$\Cn$ scattering was studied in the three-body
 $\Cx + n + n$ model. Realistic $nn$ CD Bonn potential and
a number of shallow and  deep
$n$-$\Cx$ potentials of different range  were used. All deep
potentials support deeply-bound Pauli-forbidden states
that were projected out thereby accounting for the
identity of external neutrons and those
within the $\Cx$ core in an  approximate way,
while shallow models ignore this aspect.
For all models the potential parameters were adjusted to reproduce
the experimental binding of the $\Cnn$ ground state;
most of the deep models had to be supplemented by a 3BF to achieve this goal.
Exact three-body Faddeev-type
scattering theory in the AGS version for transition operators,
extended to include also the 3BF, was implemented in the momentum-space
partial-wave framework yielding numerically accurate results for
the $n$-$\Cn$ scattering both below and above breakup threshold.

Given the weak binding of $\Cn$ and large $nn$ scattering length,
the  $\Cx + n + n$ system in the $J^\Pi=0^+$ partial wave
exhibits some features characteristic 
for Efimov physics. In particular, the presence of an
 excited $\Cnn$ Efimov state as a virtual state leads to a pole
in the $J^\Pi=0^+$  $n$-$\Cn$ effective range expansion.
The reduced effective range functions  $(1- E_k/E_0)k\cot{\delta}$ 
clearly separate for shallow and deep models, indicating the importance
of a proper treatment for  deeply-bound Pauli-forbidden states.
For some observables like the $n$-$\Cn$ singlet scattering length 
the presence of  deep Pauli-forbidden states appears to be more
decisive than the correlation with the $n$-$\Cx$ effective range.
On the other hand, the observed differences between the groups of
shallow and deep models are of comparable size as the finite
range effects found in  Ref.~\cite{shalchi:17a}, and therefore 
do not invalidate the concept of the Efimov physics 
being independent of the short-range interaction details.
However, the present work shows that deeply-bound Pauli-forbidden states
may lead to systematic shifts within the limits of finite-range 
corrections. The effect is even more important for non-observable
quantities like the expectation value of the $\Cnn$ internal kinetic 
energy.

For the elastic $n$-$\Cn$ scattering in the $J^\Pi=0^+$ partial wave
the signature of the Efimov physics, i.e., the presence of the
cross section minimum, was confirmed for both shallow and deep models.
It was also shown that without the initial antiparallel $n$-$\Cn$
polarization this minimum is, however, hidden to a large extent due to
the dominating contribution of the $J^\Pi=1^+$ partial wave.

In the hypothetical situation of  very weak $\Cn$ binding and 
near-threshold (bound or virtual) excited  $\Cnn$ state the
standard Efimovian behaviour of the  $n$-$\Cn$ scattering length and
cross section was confirmed as well, clearly
supporting Refs.~\cite{Yamashita200849,shalchi:17a,PhysRevLett.99.269201}
and excluding the possibility of near-threshold resonances predicted in 
Refs.~\cite{PhysRevC.69.061301,PhysRevLett.97.062503,PhysRevLett.99.269202}.
As both groups have solved Faddeev equations with rank-one $s$-wave potentials,
a possible explanation for this difference could be inaccurate numerical
implementation in 
Refs.~\cite{PhysRevC.69.061301,PhysRevLett.97.062503,PhysRevLett.99.269202}.

In contrast to the elastic $n$-$\Cn$ scattering, the breakup reaction
is dominated by two-body physics. The inelasticity 
parameter in the $J^\Pi=0^+$ partial wave is mostly correlated with
the ANC of the $\Cn$ bound state; this suggests a simple
$nn$-knockout picture for the reaction mechanism.

Although the present work demonstrated the importance of the
 deeply-bound Pauli-forbidden states in the low-energy 
elastic $n$-$\Cn$ scattering, further changes can be expected given
the low excitation energy of the $\Cx$ core \cite{ame2012}.
This would lead
to the $d$-wave admixture in the  $\Cn$ ground state and possibly
to $d$-wave excited states or resonances, thereby bringing 
$d$-wave corrections to the $s$-wave dominated Efimov physics of the 
 $\Cx + n + n$ system. For example, significant $d$-wave effects
have been found in the study of cold atom systems with van der Waals
interactions \cite{PhysRevA.95.032707}.

\vspace{1mm}

This work was supported by Lietuvos Mokslo Taryba
(Research Council of Lithuania) under Contract No.~MIP-094/2015
and by Alexander von Humboldt-Stiftung.
The author acknowledges also the hospitality of the Ruhr-Universit\"at Bochum
where a part of this work was performed.

\vspace{1mm}
%\begin{acknowledgments}
%\end{acknowledgments}

%\clearpage

%%%%%%%%%%%%%%%%%%%%%%%%%%%%%%%%%%%%%%%%%%%%%%%%%%%%%%%%%%%%%%%%%%%%%%%%%%%%%
%\bibliographystyle{prsty}
%\bibliographystyle{plbsty} \bibliography{abbrev,pre80,80-89,90-99,200x,clmb,ad,nreact,atomic,nd-efim,halo,dalpha} \end{document}

\end{document}